# What is the meaning of lifetime measurement?

G. Grégoire[2], C. Dedonder-Lardeux[1] and C. jouvet[1]

[1]Laboratoire de Photophysique Moléculaire du CNRS - Université Paris-Sud, Bât. 210, 91405 Orsay, France.

[2]Laboratoire de Physique des Lasers du CNRS - Université Paris-Nord, Institut Galilée, 93430 Villetaneuse, France.

The lifetime measurement of molecular excited state has been the subject of many papers and experiments. Very often the experimental data are fitted by single or bi exponential decays which in many case is the best fit that can be done owing the signal to noise ratio. The times constants obtained from these fit are often discussed in term of one species associated with one lifetime: depending on the studied system, the species can be one type of molecule, one isomer from a given molecule or local environment. How justified is this assumption?

In the last years, we have undertaken a comprehensive study of excited state lifetimes of single protonated aromatic amino acids and small peptides by means of femtosecond pump/probe photofragmentation scheme. The bottom line was to bring new insight into the non radiative processes suggested from previous studies in liquid. Since the first studies on the fluorescence properties of aromatic amino acids [1] the use of electronic energy transfer in proteins has triggered a wide variety of fundamental investigations and applications. Tryptophan has been the most studied aromatic amino acid owing to its strong fluorescence yield and the rich variety of information it provides according to the possible environments[2]. Surprisingly, the interpretation of the strong dependence of the fluorescence of tryptophan and other aromatic compounds upon local protein environment has been long and difficult. For instance, the rotamer model has been widely used to explain the nonexponential decays in tryptophan and tryptophan containing peptides [3].

In a recent experiment we have observed that both excited state of protonated tryptophan[4] and tryptophan-leucine[5] were exhibiting very similar biexponential decay . The two measured time constants were tentatively assigned to the presence of two sets of isomer. However this assumption can be questioned for these following reasons:

a) These gas phase experiments were performed at room temperature. Since these molecules are very flexible, one can seriously argue about the signification of isomers since the geometries explored by the molecule on the ground state potential energy surface should be very wide.

b) In protonated tryptophan, the short observed lifetimes (400 fs, 15 ps) result from the coupling between the initially excited $\pi\pi^*$ state and the $\pi\sigma^*$ state, dissociative along the NH coordinate of the amino group. Following the SDDJ model [6] the energy gap between the $\pi\sigma^*$ and the $\pi\pi^*$ states is the key factor which determines the excited state lifetime. This model can also explain the longer lifetime observed in tyrosine as compared to tryptophan: in tryptophan the lifetime is shorter because the energy gap between $\pi\pi^*$ and $\pi\sigma^*$ excited states is smaller than in tyrosine. Recent ab-initio calculations on excited states of protonated tryptophan [7] have shown that this energy gap is strongly dependent upon the conformation of the molecule. Since many conformers are populated at room temperature, we should expect experimental transients more complex than a biexponential decay.

As a test we will make the assumption that a molecular system is composed of isomers described by different distributions. Three simple distributions of lifetimes have been tried: two gaussian distributions centred at 100 a.u. with a time width of 5 and 50 a.u. and a square distribution from 0.1 to 300 a.u. The resulting observed transient is the sum of all the exponential decay of individual isomer integrated from 0 to 300.

In figure 1, 2 and 3 are plotted the curves from 0 to 300 a.u. obtained by numerical integration of the two Gaussian distributions and the square distribution reported in the bottom of each graph. The resulting decay have been fitted by a single, double and a triple exponential decay respectively and plotted with open circle.

The fitting function used is:
$Y(t) = y_0 + A_1 \exp(-x/t_1) + A_2 \exp(-x/t_2) + A_3 \exp(-x/t_3)$
with the $A_i/t_i$ parameters reported on each graph.

As it can be seen, the fits are almost perfect with a correlation coefficient $R^2 = 1$ independently of the initial lifetime distribution. In most of the experiments, owing the signal to noise ratio, It will be considered as very good fit. From that, one should deduce that the observed transients can be unambiguously assigned to respectively one, two and three sets of conformers, which is obviously wrong in the last two cases.

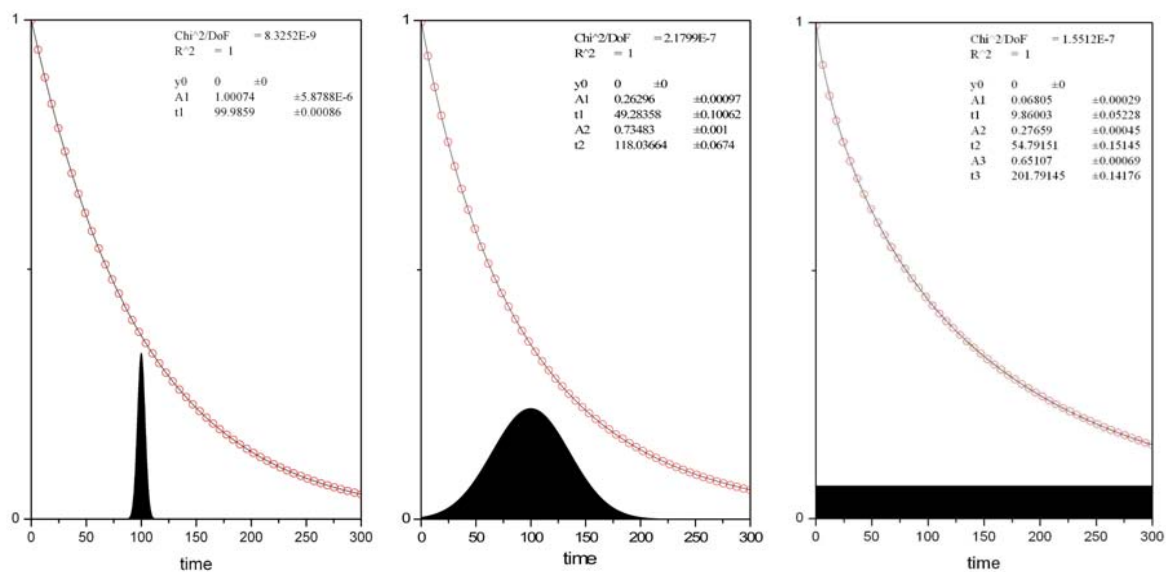

*Figure 1 : Calculated decay with the distribution presented in Black and the multiexponential fit.*

In fact this very simple simulation shows that lifetime measurement (unless extremely good signal to noise ratio is achieved) cannot differentiate between a molecular system in which a set of molecules have distinct lifetimes and a system where molecules have all possible lifetime with equal probability.

In the case of the protonated tryptophan[4], the fact than we can control the fragmentation pathway with femtosecond pulses at very short time suggests that the isomer hypothesis is may be still relevant.

Conclusions

This paper is not the only one in which a strong revision of the conclusions derived from the observation of bi exponential decays is needed.

References.